*Article*

# Aerobars Position Effect: What is the Interaction Between Aerodynamic Drag and Power Production?

**TEROL Sébastien\*[1,2], COSTES Antony[2], MALMERT Alexandre[2], BRUNET Emmanuel[3], SOTO-ROMERO Georges[1]**

[1]  LAAS-CNRS, Université de Toulouse, CNRS, INSA, Toulouse, France ; sterol@laas.fr ; gsotorom@laas.fr
[2]  Alten Labs, 7 Rue Alain Fournier, 31300 Toulouse, France ; antony.costes@alten.com ; alexandre.malmert@alten.com
[3]  French Cycling Federation, 78180, Saint-Quentin en Yvelines, France ; e.brunet@ffc.fr

    \*  Correspondence: (ST) sterol@laas.fr.
    **Author version**

**Abstract:**

    Extensive research has been dedicated to optimizing the cyclist's position on the bike to enhance aerodynamic performance. This study aims to further investigate the aerobars position effect on cycling speed. Drawing from previous work (Fintelman et al., 2015), a relationship is established between position variations and hip angle, a critical determinant of power output. Based on a 3D scan of an elite athlete on his Time Trial (TT) bike, a digital twin with upper body mobility is created, utilizing inverse kinematics with aerobars as a root. Adjustments to the aerobars position translate into alterations in the cyclist's upper body posture. These changes influence both aerodynamic drag -quantified by Computational Fluid Dynamics method (CFD)- and hip angle -computed by 3D software, directly affecting the athlete's capacity for power generation. The interplay between aerodynamic efficiency and power output is analyzed, with varying parameters such as speed and slope angle considered to ascertain the optimal aerobar position for individual athletes in a specific cycling context. Results show impactful variations in cycling speed as a function of the aerobars position, the latter having a strong influence on aerodynamic drag and theoretical power production.

**Keywords:**

Aerodynamic drag; Power Production; Computational Fluid Dynamics; CFD; Inverse Kinematics; Aerobars.

## 1. Introduction

Aerodynamic and energy optimization in the field of sports represent an essential research area that directly influences the performance of athletes and sports equipment (Crouch et al., 2017; Malizia & Blocken, 2021). These topics aims to maximize the efficiency of movements by minimizing aerodynamic resistance and optimizing energy use. Its significance is crucial in a context where winning margins often come down to fractions of a second or millimeters.

The fundamental objective of this work is to comprehend how aerodynamic and energy efficiency principles can be simultaneously applied to enhance the performance of cyclists. By redefining the position of the aerobars, both aerodynamic drag and cyclist power are impacted (Faulkner & Jobling, 2020), leading to an optimal position that reduces aerodynamic drag and optimizes the energy produced by the cyclist .

The research field concerning aerodynamic and energy optimization in sports is constantly evolving, driven by advancements in numerical modeling, Computational Fluid Dynamics (CFD), and data analysis techniques. Noteworthy works include those of (Grappe et al., 1997), focusing on optimizing cyclists' aerodynamic position, as well as those of (Blocken et al., 2018), proposing a numerical modeling of airflows surrounding moving athletes.

The article's approach is to automate this process using digital tools to modify the aerobars position.

## 2. Materials and Methods

In order to achieve the pre-established objectives, a 3D scan of a professional athlete was performed with the Freestyle 3D scanner (Faro, Lake Mary, USA). Based on this scan done in a time trial position, the aim was to animate the geometry derived from the scan to digitally modify the athlete's position.

To carry out the geometry animation, the animation software Blender (Blender Foundation, Amsterdam, Netherlands) was used. The animation method involves creating a human skeleton using different bones (Error! Reference source not found.) to complete a 3D ringing (Schaffarczyk et al., 2022). Once the skeleton is created, by using inverse kinematics, it is possible to modify the position of the entire athlete simply by moving the control bones joined with the aerobars.

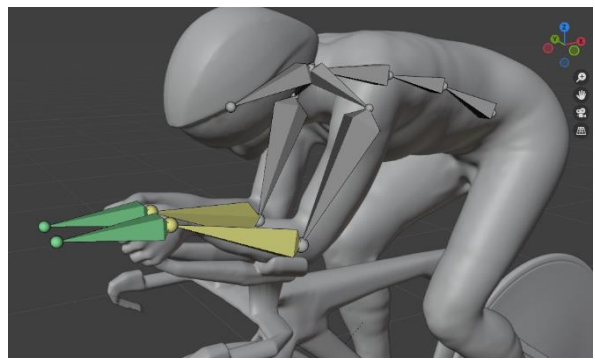

**Figure 1.** Blender Screenshot. This figure shows the armature of the geometry based on the 3D scan of a professional athlete.

From this skeleton model created in Blender, it is possible to automate the movement and generation of geometries in the STL format through the Blender to Python API. This automation is carried out by considering several limit positions corresponding to the limits imposed by the

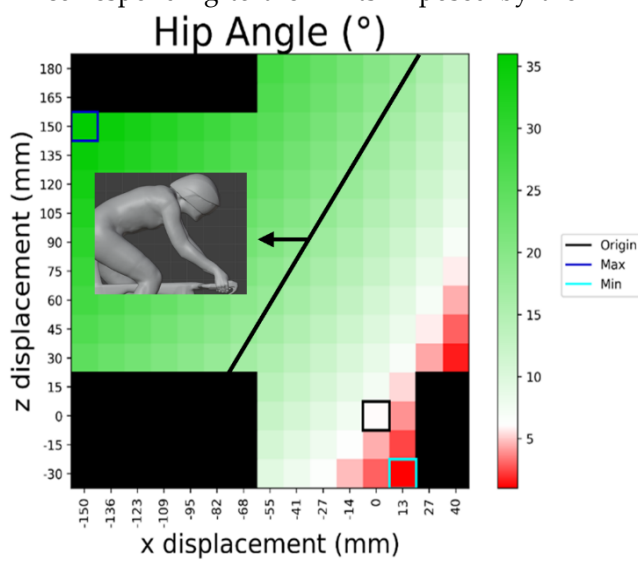

**Figure 2.** Hip Angle evolution according to the armature displacement in the zOx 2D plan.

*The hip angle corresponds to the angle between the greater trochanter and the acromion.*

*Left part of the diagonal line indicates a torso angle greater than the one at the hoods position.*

*"Origin" corresponds to the initial 3D scan position, "Min" to the lowest value and "Max" to the highest value.*

Union Cyclist International (UCI) (*Règlement UCI du cyclisme sur route*, 2023) or simply anatomical limits involving limb collisions. For each position, a hip angle is obtained by reading the angle between the horizontal and the vector created by the armature near greater trochanter and acromion position.

The black line on the **Figure 2** correspond to an isoangle of 16.6°. The value of 16.6° for the hip angle correspond to an upright position with the hands on the hoods. Every position with a higher hip angle should be considered as the common upright position, hand on the hoods.

remains the same in the case of our athlete, we use $a = 2.5\ W/°$. If the athlete develops $300W$ at an initial hip angle of 5.925°, we can determine $b = 285\ W$ by solving:

$$f(5.925) = 300 = 2.5 * 5.925 + b \quad (1)$$

Hence the final form of the equation:

$$f(x) = 2.5 * x + 285 \quad (2)$$

This athlete-specific equation is the prerequisite for the implementation of the analysis methodology. It needs individual testing for accurate outputs.

The CFD part is carried out using the open-source software OpenFOAM (Open-source Field Operation And Manipulation) in

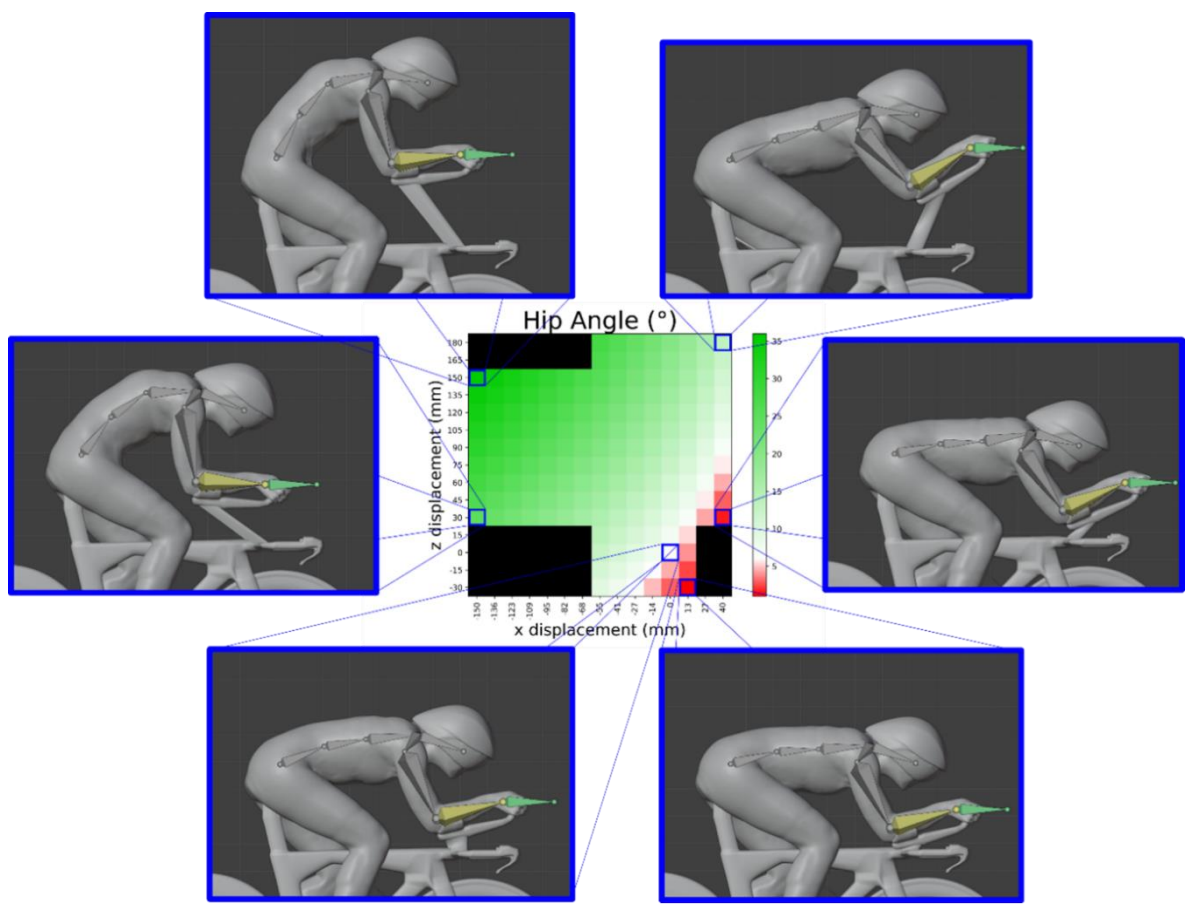

**Figure 3.** Relation between heat map format and corresponding position
*Each square of the heat map corresponds to a position of the 3D geometry. This figure illustrates the change in hip angle by associating the Blender screenshot with the corresponding position.*

According to previous literature results, particularly those of (Fintelman et al., 2015), it is possible to link the hip angle of our geometry with the maximum power developed by the athlete during an incremental test. The experimental values presented in the article have been approximated by an affine function $f(x) = ax + b$. Considering that the evolution its 2312 version. The geometry automation allows us to generate the CFD simulation folders. Initially, the study focuses on a time efficient approach with the k-omega SST turbulence model (OpenCFD Ltd., 2016; Rumsey, 2021). The total number of cells for one simulation amounts to one million. The selection of a coarse mesh from a grid convergence study is made to prioritize computational efficiency.

The aim of the study is to test the various positions generated by the python code by linking the calculated power output from hip angle at each iteration with the aerodynamic drag provided by the CFD simulations. Then, the algorithm will be able to suggest position recommendations according to various road slope angles. Slope values ranging from -5% to +5% in steps of 1% were tested.

A velocity prediction program is linked to the algorithm in order to define a cycling speed for each system configuration. The resistive powers applied to our model consist of the power required to overcome the aerodynamic drag ($P_{aero}$), rolling resistance ($P_{roll}$) and gravity related to the gradient encountered ($P_{grad}$). By carrying out a power balance applied to the cyclist using equation (3), we can set the speed at which the resistive powers are equal to power output for each position. This speed corresponds to the maximum speed the cyclist can achieve with one position and associated power output (**Figure 5**).

$$P_{output} = P_{aero} + P_{roll} + P_{grad} \qquad (3)$$

## 3. Results

### 3.1. Power Output Analysis

The values from hip angle (**Figure 2**) paired with equation (2), allow the model to link hip angle and power output. Series of values are obtained defining the power production as a function of aerobars position (**Figure 4**). This figure highlights the fact that with current inputs, the higher the rider's hip

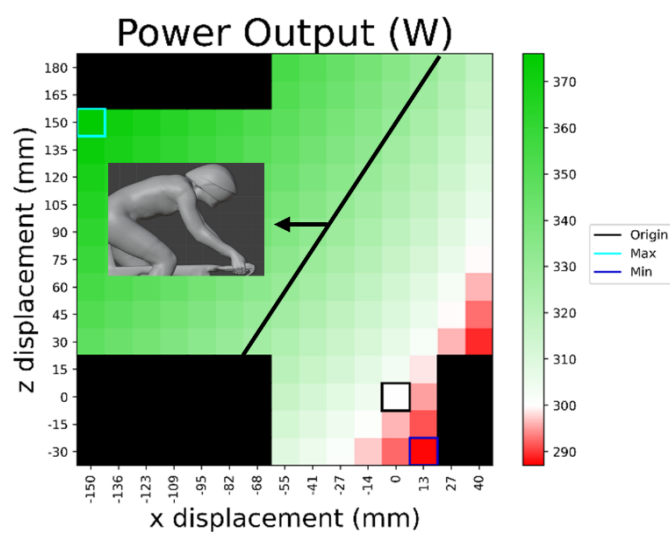

**Figure 4.** Power Output evolution according to the armature displacement in the zOx 2D plan.
*The power output is directly linked with the hip angle by the equation from Fintelman et al. (2015).*

angle, the higher power production is and vice versa.

### 3.2. CFD Analysis

Series of simulations were carried out with different geometry derived from the animation of our initial 3D model. A connection can be established between the aerobars position modification and the hip angle (**Figure 2** & **Figure 3**). On the one hand, the further the aerobars are moved forward and downward, the more the hip angle closes. The rider is then in Time Trial (TT) position. On the other hand, when the hip angle goes over 17 degrees, corresponding to an upright position with the hands on the hoods, the position is considered as an upright position.

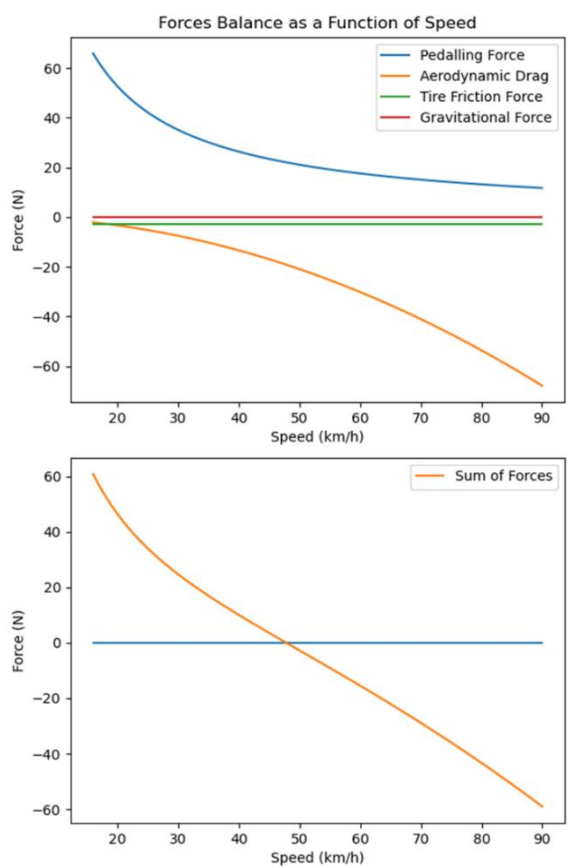

**Figure 5.** (A) Forces applied to the model
(B) Sum of forces analysis

Similarly to **Figure 4**, a relationship can be drawn between the aerobars position modification and the aerodynamic drag calculated from CFD simulations (**Figure 6**). A displacement in the negative x axis and positive z axis induces an upright position leading to a high drag value. On the other hand, a displacement in the positive x axis the hoods. On the other hand, on a high negative gradient, in freewheel situation the aerodynamic impact greater, and the power output required to maintain speed is low. An aerodynamic position with aerobars makes sense in this case.

With the data input of the study, a compromise between an aerodynamic position and a power production efficient position is suggested for each slope angle. **Figure 7** show the example of two different slope angle where the model suggests an optimal position. When the gradient is positive by 4%, **Figure 7B** indicates an

## Drag compared to baseline (%)

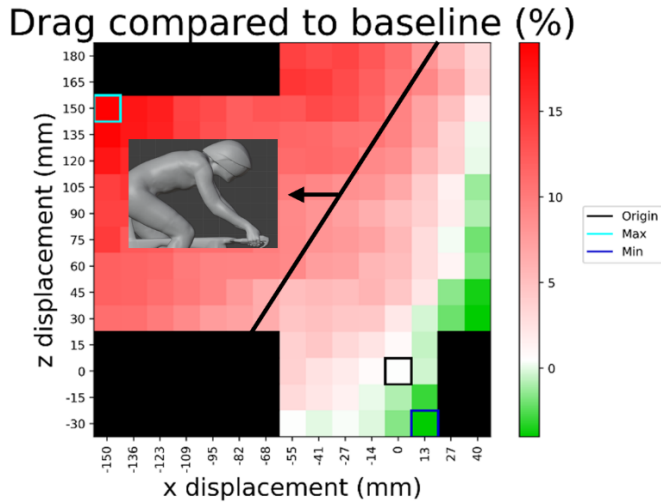

**Figure 6.** Aerodynamic drag evolution according to the armature displacement in the zOx 2D plan.
*The drag value at the origin position is set as baseline for other values. The red colors corresponding to less aerodynamic positions have a higher drag value. The green color show new position that are more aerodynamic than the original one.*

and negative z axis correspond to an aerodynamic position of the cyclist leading to a 17.4% lower drag than the upright position, hands on the hoods, at a TT speed.

Comparing **Figure 4** and **Figure 6** introduces the core of the current study. As hip angle increases, power production rises, and aerodynamic drag increases accordingly. There is therefore a compromise to be found in order to optimize the rider's position for given road conditions.

### 3.3. Velocity Prediction Program According to Slope Angle

The second part of the study focuses on the effect of slope on position choice. On the one hand, high positive slope in climbing situation involves lower efficiency and a need of high-power output from the cyclist to maintain even a low speed. Moreover, at low speed, aerodynamics has a considerably reduced impact so the recommended position will be an upright position, hands on

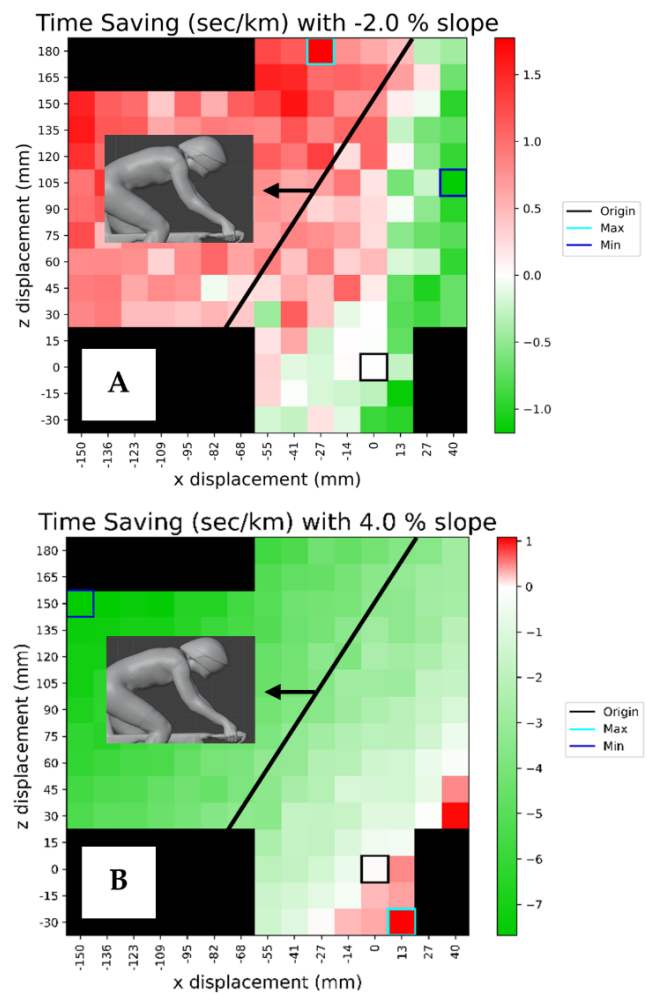

**Figure 7.** Potential time saving according to different slope angles.
*A positive value of the heat map means a loss of time compared to baseline.*
*A negative value means a potential time saved compared to baseline.*

    A.    -2% positive angle
    B.    4%, negative angle

Upright Position, whereas **Figure 7**A, corresponding to a negative 2% slope angle, indicates a new Aerodynamic position that should save about 1 second per kilometer.

According to the study, there is an optimum position for every slope angle. To illustrate this result, considering 6 random positions (**Figure 8**). The model can determine the Optimal position for every slope angle **Figure 9**.

different positions that are supposed to be optimal. Near to the freewheel zone, the optimal position suggested is the most aerodynamic position whereas near to the climbing zone the position allowing the best power output is needed. In our example, the algorithm predicts a 1.5 second per kilometers time saved for a -4% slope angle compared to the 3D scan baseline position.

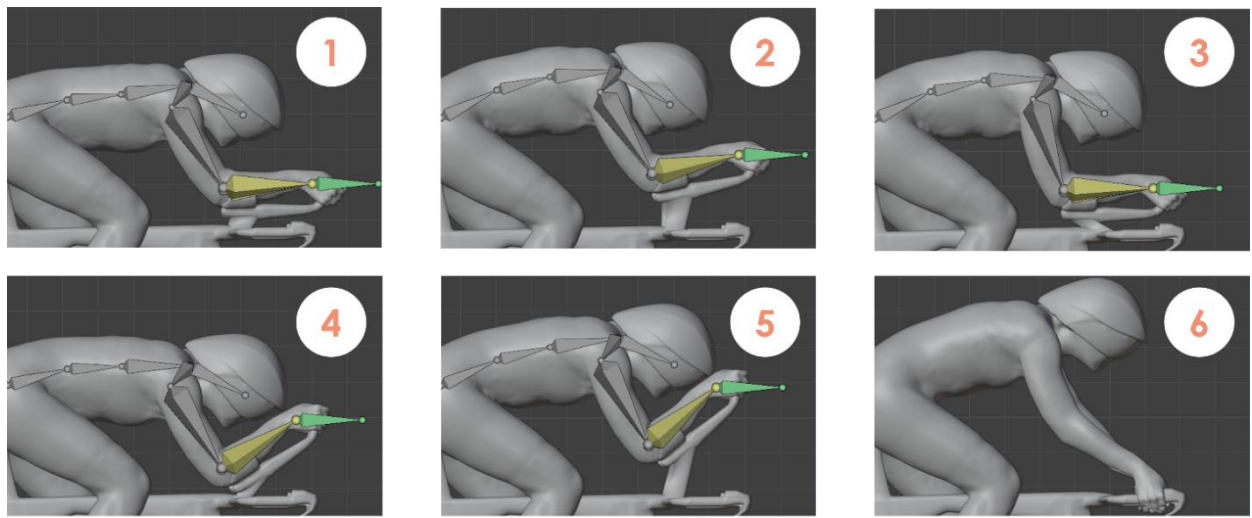

**Figure 8.** Six random positions from the 3D scan and animated using Blender software.

According to the recommendation of the algorithm, from the freewheel zone, fixed here at -4% slope angle, to the climbing zone, fixed at +4% slope angle, there are only 4

For a slope angle of +4%, the upright position leads to a 3.6 second per kilometers time saved compared to baseline position. Between freewheel and climbing zones, the position recommended by the model represents the best compromise between an aerodynamic position and a power efficient position.

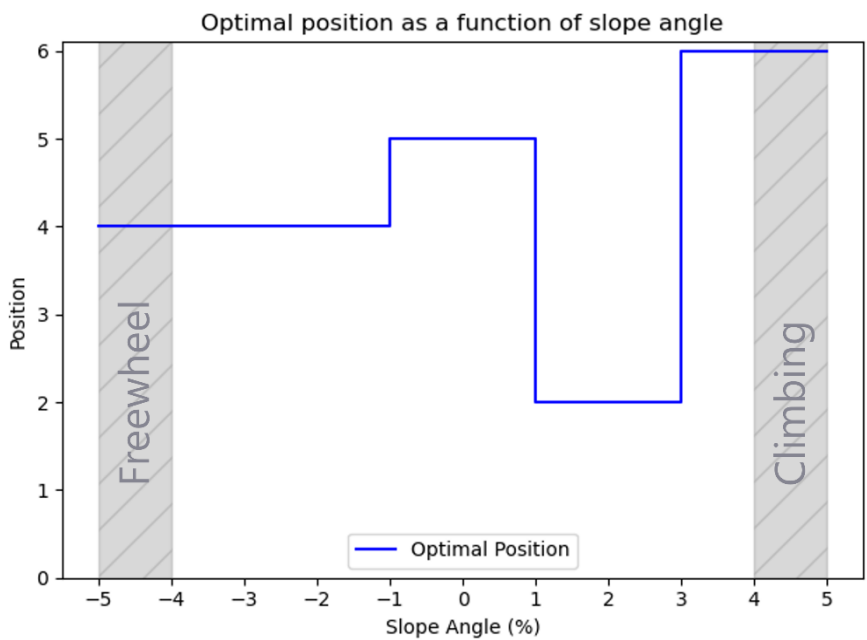

**Figure 9.** Optimal position suggested by the model among the 6 random positions of Figure 1 according to the slope angle.

## 4. Discussion

The determination of an optimal position according to several road conditions leads to potential time saving for a specific athlete. This optimization process requires power output and aerodynamic data of a cyclist. The power output data come from Fintelman et al. 2015 methodology using affine function to link hip angle to power output of a given athlete. The aerodynamic data come from CFD simulations with a time efficient model in order to quickly explore a wide position range. Many studies have been carried out analyzing the aerodynamic effects of different cyclist positions with experimental measurement (Grappe et al., 1997), in wind tunnel or in CFD (Defraeye et al., 2010). The drag analysis from **Figure 6** shows a correlation with literature results. Indeed, in an upright position, with a negative displacement in the x axis and a positive displacement in the z axis, the drag is 17.4 % higher than with a positive displacement in the x axis and a negative displacement in the z axis corresponding to the most aerodynamic position.

From aerodynamic drag and power output data, forces applied to the cyclist is analyzed (equation (3)). This analysis methodology has been developed and is validated (Martin et al., 1998). To further enhance this model, slope angle is tested in the present study. The slope angle modification analysis shows important findings on the position choices. Negative gradients suggest more aerodynamic positions. For positive gradients, the predicted time saved evolves rapidly, underlining the importance of not staying on the aerobars when the gradient is too steep. There is therefore a slope value at which it is necessary to switch from an aerodynamic position to an upright position that generates more power. More generally, there is an optimal position for each slope angles.

This study leads to promising results suggesting an optimal position for one given cyclist according to different road condition such as slope angle. These suggestions are strongly linked with the input data quality. Indeed, better input data involve better position recommendation. Power production tests can be performed with incremental tests on a cycle ergometer and adapted protocols depending on expected race situations. The power output as a function of hip angle could also be modified with training protocols. Power data have to be individualized, corresponding to the same athlete as the 3D scan, allowing the model to accurately use the aerodynamic data from CFD. The time efficient CFD model used in this article can be replaced by a higher fidelity model for more accurate results at the price of higher computational times. This high-fidelity model could include finer mesh but also dynamic implementation such as the cyclist's pedaling motion (Griffith et al., 2019; Javadi, 2022). The method proposed in this present paper should be adapted to a specific cycling route with adapted geometry, data and road profile (Schaffarczyk et al., 2022). Additional parameters can be analyzed by implementing new degree of freedom such as wind speed. Aerobars geometry modification due to ringing displacement can be optimized. A displacement of the aerobars along the y axis can be implemented in the Blender model as well as an angle modification over the head, leading to a multiplication of positions to be processed. This accumulation of simulations can be compensated by using surrogate models to reduce the total computational time. Future work should focus on experimental validation using wind tunnel testing or field conditions and determination of power output for different hip angles in laboratory or field conditions.

## 6. Conclusions

This article presents a complete analysis of aerobars position modification effects on cycling speed. Based on a 3D scan of a professional athlete, a digital twin was created and animated using the Blender software. This model, animated at the aerobars, was used to generate a multitude of static geometries with different cyclist positions, ranging from an aerodynamic position to an upright position with hands on the hoods. Each position is associated with a hip angle. Using the results from Fintelman et al. 2015, this hip angle can be related to a given power output for a rider. The range of displacement of the aerobars was determined by considering the regulations imposed by the UCI and limb collisions. For each position, an aerodynamic study was carried out to analyze the impact of aerodynamic drag on a given position. By carrying a force analysis on the cyclist, it is possible to determine which position is the most suitable for a given configuration. By implementing slope angle as a new variable parameter to the study, the model provides conclusions regarding the position to adopt according to the gradient. On a downhill road, at the limit of the freewheel zone (-4%), the model indicates that the most aerodynamic position is more suitable, while on an uphill road near the climbing zone (+4%) an upright position that generates maximum power is preferred. Between -4% and +4%, the optimum position is a compromise between aerodynamic and power production. Based on these results, the model predicts that it makes sense to change position for different slopes. Further work using this methodology should focus on applying it on a full race profile including wind conditions to design new aerobars involving the most efficient cyclist position for each different racecourse and cyclist characteristics.

**Acknowledgments:** I would like to express my special thanks to the athlete who was digitalized using a 3D scan method.

**Conflicts of Interest:** The authors declare no conflict of interest.